# Construction automatique d'ontologie et enrichissement à partir de ressources externes

Eric KERGOSIEN
UPPA, Laboratoire LIUPPA,
Département Informatique,
64000 PAU
eric.kergosien@univ-pau.fr

Marie-Noelle BESSAGNET
UPPA, Laboratoire LIUPPA, IAE,
Avenue du doyen Poplawski,
64012 PAU
marie-noelle.bessagnet@univ-pau.fr

Mouna KAMEL
IRIT - UPS
118 Route de Narbonne
31 068 Toulouse
kamel@irit.fr

Nathalie AUSSENAC
IRIT - UPS
118 Route de Narbonne
31 068 Toulouse
Aussenac@irit.fr

Christian SALLABERRY
UPPA, Laboratoire LIUPPA, FDEG,
Avenue du doyen Poplawski,
64012 PAU
christian.sallaberry@univ-pau.fr

Mauro GAIO
UPPA, Laboratoire LIUPPA,
Département Informatique,
64000 PAU
mauro.gaio@univ-pau.fr

## ABSTRACT

La construction automatique d'ontologies à partir de textes est généralement basée sur le texte proprement dit, et le domaine décrit est circonscrit au contenu du texte. Afin de concevoir des ontologies sémantiquement plus riches, nous proposons d'étendre les méthodes classiques en matière de construction d'ontologie (1) en prenant en compte le texte du point de vue de sa structure et de son contenu pour construire un premier noyau d'ontologie, et (2) en enrichissant l'ontologie obtenue en exploitant des ressources externes (textes grand public et vocabulaires contrôlés du même domaine). Ce papier décrit comment ces différentes ressources sont analysées et exploitées. Nous avons appliqué cette méthode sur des textes géographiques et avons évalué le bénéfice induit par une ontologie plus riche (par rapport à une première taxonomie du domaine) dans le contexte du typage des entités nommées spatiales. Les résultats ont été améliorés de façon significative.

## Categories and Subject Descriptors

H.3.1 **Content Analysis and Indexing**: *Abstracting methods - Dictionaries - Indexing methods - Linguistic processing - Thesauruses.*

## 1. INTRODUCTION

Ce travail a pour point de départ une chaîne d'annotation automatique d'entités nommées (EN) basée sur une analyse lexicale et syntaxique de fonds documentaires textuels. Cette chaine nous permet d'interpréter des EN candidates selon un typage parfois peu satisfaisant. A cet effet, nous présentons une démarche dont le principal objectif est d'améliorer ce typage afin de mieux rapprocher et/ou distinguer deux EN introduites différemment dans le texte. Par exemple, les syntagmes nominaux *le lac d'Artouste*, et *le mont d'Artouste*, ou encore, *le président Mitterrand* et *les années Mitterrand* doivent être typés différemment en vue d'une représentation adaptée.

La démarche globale que nous proposons consiste à concevoir, puis à enrichir une ontologie de domaine. Cette approche est relativement classique en ce qui concerne l'annotation de documents. La spécificité de ce travail est liée aux ressources utilisées pour la validation et le calcul de représentations associées aux EN détectées. L'accès optimisé à ces ressources est une priorité ; par conséquent, l'ontologie que nous allons bâtir doit également refléter la structuration de ces ressources.

Ainsi, la démarche retenue se veut plus spécifique et consiste tout d'abord à créer une ontologie à partir de documents de description de ressources. L'intégration d'une telle ontologie dans notre chaine d'annotation nous permet un typage automatique de 10% des EN distinctes (soient 33% des occurrences d'EN). Enfin, l'enrichissement de cette ontologie à partir de ressources sémantiques structurées externes nous mène à un typage attendu de 50% des EN distinctes (soient 75% des occurrences).

Ce papier se compose de trois parties principales. La première partie présente les principales méthodes utilisées en matière de construction d'ontologies à partir de textes constituant un corpus. Notre objectif étant d'obtenir une ontologie riche offrant une couverture sémantique étendue d'un domaine cible, nous présentons dans la deuxième partie comment l'exploitation de ressources externes telles que des textes grand public ou d'autres ressources structurées ou peu structurées peuvent contribuer à enrichir l'ontologie. La troisième et dernière partie décrit une application de notre approche dans le cadre du projet ANR GEONTO qui vise le domaine spécifique des ontologies géographiques. Une évaluation de notre approche est également présentée dans cette partie.

## 2. TRAVAUX CONNEXES

Les méthodes de construction d'ontologies à partir de textes privilégient souvent l'analyse du texte proprement dit, que ce soit selon une approche statistique ou linguistique (Nédellec et Nazarenko, 2003), (Aussenac et al., 2008), (Maedche, 2002), (Buitelaar et al., 2005). La plupart de ces travaux montrent la nécessité d'intégrer différents outils de TAL, et soulignent la complémentarité entre identification de concepts et extraction de relations. L'étape d'enrichissement d'une ontologie consiste alors à découvrir de nouveaux concepts, de nouveaux termes, et/ou de nouvelles relations entre concepts. Pour cela, deux familles de techniques d'identification de concepts et de relations existent : les approches statistiques et les approches linguistiques.

Les approches statistiques consistent à étudier généralement les termes co-occurrents par analyse de leur distribution dans le corpus (Agirre et al., 2000) ou encore par des mesures calculant la probabilité d'occurrences d'un ensemble de termes (Velardi et al., 2001), (Neshatian & Hejazi, 2004). Les relations peuvent être identifiées par des calculs de similarité entre leurs contextes syntaxiques (Hindle, 1990), (Grefenstette, 1994), par prédiction à l'aide de réseaux bayésiens (Weissenbacher et Nazarenko, 2007) ou de techniques de Text Mining (Grcar et al., 2007), ou encore par inférence de connaissances à l'aide d'algorithmes d'apprentissage (Guiliano et al., 2006). Ces méthodes sont efficaces, mais elles nécessitent une intervention humaine fastidieuse pour le positionnement des concepts dans l'ontologie, ou n'identifient pas toujours la sémantique de la relation.

L'approche linguistique peut être utilisée pour découvrir des concepts via des règles d'association, mais est plus largement mise en application dans le cadre d'identification de relations. Dans ce cas, elle fait appel à des analyses syntaxiques ou des calculs de dépendance pour identifier les relations argumentatives (sujet, verbe, objet) (Jacquemin, 1997), (Bourigault, 2002), ou définit des patrons lexico-syntaxiques pour reconnaître les marques linguistiques des relations sémantiques (Aussenac et Seguela, 2000). Ainsi la sémantique des relations est bien identifiée, mais la variabilité de leur sémantique et de leur expression en corpus oblige à multiplier les patrons et rend l'approche coûteuse.

Ces techniques s'appliquent au niveau interne de la phrase, alors que d'autres études ont pour niveau d'analyse le texte lui-même. L'objectif est alors assez différent : il ne s'agit plus de trouver des relations entre concepts, mais des relations sémantiques plus diffuses entre les différentes unités textuelles repérées. Ces liens peuvent être décelés soit à l'aide de marqueurs linguistiques (Asher et al., 2001), soit en exploitant la structure matérielle du texte, (Virbel & Luc, 2001) ayant montré que la matérialité d'un texte participe à son sens, soit en combinant les marqueurs linguistiques et la structure du texte (Charolles, 1997).

Nous proposons dans ce travail d'étendre ces techniques pour pouvoir construire des ontologies plus riches dans la mesure où elles seront d'une part construites à partir de documents pour lesquels non seulement le contenu sera pris en compte, mais également sa structure matérielle, et qui d'autre part seront enrichies à l'aide de ressources externes structurées ou peu. Ces ontologies refléteront alors différents points de vue, permettant ainsi un typage d'informations plus performant que ne le ferait une ontologie issue d'un texte ou d'un corpus.

## 3. DEMARCHE GENERALE

L'approche que nous proposons consiste à produire une ontologie à partir d'un document, en analysant à la fois sa structure matérielle et son contenu pour obtenir une ontologie plus riche que ne fournirait une de ces analyses prises indépendamment. Afin d'améliorer le typage d'EN distinctes, l'idée est d'enrichir cette ontologie en découvrant essentiellement de nouveaux concepts dans des ressources externes.

## 3.1 Construction d'une ontologie à partir de description de documents spécifiques

Nous proposons ici d'étendre les approches classiques d'identification de relations en nous appuyant sur la structure matérielle du texte et son contenu. En effet, des moyens de mise en forme du texte (titres de sections et de sous-sections, énumérations, définitions, etc.) caractérisent essentiellement des relations hiérarchiques (Jacques, 2005). Lorsque le document, destiné à un support numérique, adopte un format tel que HTML, SGML, XML, etc., la structure du document est facilement repérable. Nous nous appuyons sur la caractérisation de ce type d'élément pour construire un premier noyau d'ontologie. L'analyse du texte permettra ensuite d'enrichir ce noyau d'ontologie.

### 3.1.1 Analyse de la structure du document

Cette étude tient compte de documents au format XML. Nous considérons dans ce cas que lorsque chaque syntagme marqué par des balises réfère à un seul concept, alors la hiérarchie des balises traduit des relations sémantiques. Nous basons notre analyse sur la règle suivante décrite figure 1.

> Lorsque  - A et B sont des balises qui sont sous la portée d'une même balise O
> - $C_1$ et $C_2$ sont des concepts respectivement étiquetés par les unités textuelles marquées par A et B
> Alors  une relation sémantique existe entre $C_1$ et $C_2$.

**Figure 1. Règle d'extraction de relation sémantique**

L'instanciation de cette règle nécessite l'intervention d'un expert ; l'analyse automatique du corpus à partir de ces règles fournit le noyau d'ontologie. Nous illustrons nos propos à l'aide de l'extrait de fichier XML décrit figure 2.

```
<class>
    <className> Oronyme </className>
    <valueName> Grotte </valueName>
</class>
```

**Figure 2. Extrait d'un fichier XML**

L'analyse de l'expert conclut à une relation d'hyponymie entre les unités textuelles marquées par les balises <className> et <valueName>, balises qui sont sous la portée de la balise <class>. Le concept *Grotte* est créé en tant que concept fils du concept *Oronyme*.

### 3.1.2 Analyse du texte

Le corps du document XML correspond au texte en langage naturel et peut contenir de l'information intéressante à exploiter pour enrichir l'ontologie obtenue à l'issue de l'exploitation de sa structure (section 3.1.1). Selon C. Barrière (Barrière et Agbado, 2006) et Hearst (Hearst, 1992), un des moyens de qualifier "des contextes riches en connaissances" est qu'ils contiennent des marques linguistiques de relations sémantiques. Nous avons choisi d'utiliser des patrons lexico-syntaxiques pour repérer des relations sémantiques (Auger et Barrière, 2008). Un patron lexico-syntaxique décrit une expression régulière, formée de mots, de catégories grammaticales ou sémantiques, et de symboles, visant à identifier des fragments de texte répondant à ce format. L'application de tels patrons nécessite de traiter préalablement le texte en appliquant différents outils du TAL (tokenizer, lemmatiseur, analyseur syntaxique, etc.). Les patrons exploitent les étiquettes morpho-syntaxiques ou sémantiques attribuées par ces logiciels. Ainsi, la forme des patrons dépend à la fois du logiciel de définition et de projection des patrons, et des analyses et étiquetages effectués sur les textes.

L'approche par patrons complète l'exploitation de la structure des documents en identifiant de nouveaux concepts, de nouvelles relations, voire des propriétés dans les parties rédigées du document.

Prenons pour exemple le patron lexico-syntaxique suivant dont le rôle est d'identifier et d'annoter les relations d'hyponymie dans le texte. Ce patron est formulé selon la syntaxe JAPE[1] (figure 3).

```
({Terme}
{Token.lemma=="est"} {Token.lemma="un"}
{Terme}) : annot
- - >
:annot.HYPONYMIE = { kind= "est-un", rule = R0}
```

**Figure 3. Définition d'un patron lexico-syntaxique pour identifier la relation d'hyponymie**

Lorsqu'un terme est suivi du syntagme verbal "*est un*", lui-même suivi d'un terme, alors l'ensemble est annoté *HYPONYMIE*, et un traitement spécifique peut alors être associé.

## 3.2 Enrichissement d'ontologie à partir de textes grand public

L'ontologie générée selon la méthode présentée §3.1 apporte un premier élément de réponse à notre volonté de proposer au grand public un accès structuré aux ressources d'un domaine cible. Cependant, afin d'assurer une couverture sémantique du domaine cible la plus importante possible, nous proposons une méthodologie permettant d'enrichir l'ontologie à partir d'un échantillon de documents textes grand public représentatif du domaine cible. Ce choix s'impose notamment par le fait que nous ne souhaitons pas dénaturer le domaine d'application. La méthodologie proposée (figure 4) se décompose de la façon suivante : (i) identification et validation des EN dans le texte ; (ii) identification des termes liés aux EN non présents dans l'ontologie présentée §3.1 ; (iii) enrichissement de l'ontologie.

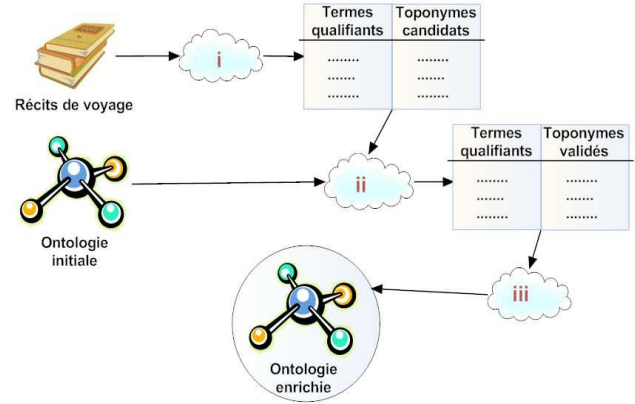

**Figure 4. Démarche d'enrichissement d'ontologie**

(i)Identification des EN dans le texte.

La reconnaissance d'EN (Named Entity Recognition (NER)) consiste à assigner un mot ou un groupe de mots à un ensemble de types d'entités standards à extraire : organisation, personne, lieu, date, heure, monnaie et pourcentage, par exemple. Le moyen le plus simple d'identifier des EN consiste à comparer le contenu d'un document à des listes de noms préétablies. Une autre approche consiste à analyser davantage le contexte via des règles de grammaires. Ces règles peuvent être générées manuellement (représentations des connaissances de l'expert) ou bien automatiquement à l'aide de techniques issues des méthodes d'apprentissage. Des méthodes de reconnaissance d'EN plus sophistiquées utilisent à la fois des listes et des règles de grammaires, complémentarités que nous avons mises en place dans notre chaîne de traitement.

(ii)Identification des termes associés aux EN

Une chaine de traitement automatique de la langue (TAL) est généralement composée de quatre principaux modules (abolhassani, 2003) : (1) la « tokenisation » qui divise le document en unités lexicales ; (2) l'analyse lexicale et morphologique qui comprend la reconnaissance et la transformation des unités en lexemes ; (3) l'analyse syntaxique et l'analyse des dépendances grammaticales qui supporte l'étiquetage grammatical (nom commun, adjectif, adverbe, etc.) des différentes unités ; (4) l'analyse sémantique qui intègre des informations contextuelles afin d'interpréter et de constituer de nouveaux lexèmes porteurs de sens.

La reconnaissance d'EN est généralement réalisée dans le module (2) à l'aide de règles typographiques et lexicales : selon le domaine, des listes de termes introducteurs d'EN peuvent être associées à ces règles, ou encore des dictionnaires d'EN peuvent être utilisés.

L'identification de termes ou de groupes de termes associés aux EN peut avoir un intérêt important dans la phase de typage/désambigüisation de l'EN. Les modules (3) et (4) d'une chaine de TAL permettent le marquage de syntagmes nominaux liés aux EN.

(iii)Enrichissement de l'ontologie de domaine

---

Nous proposons d'enrichir l'ontologie de domaine en nous appuyant sur ces termes associés aux EN ou sur une ressource de type vocabulaire contrôlé (thésaurus notamment). Les thésaurus sont des vocabulaires contrôlés de termes gérant des relations hiérarchiques, associatives et d'équivalence. L'une des caractéristiques d'un vocabulaire contrôlé est que chaque terme le constituant a un seul sens et que ce sens est représenté par un seul terme. Chaque terme portant un sens, que nous appelons vedette, peut être relié à des termes qui auront le même sens dans le thésaurus, termes appelés « termes rejetés » ou « termes employés pour ». Lorsque le thésaurus est utilisé par les catalogueurs pour indexer les documents, ceux-ci doivent utiliser le terme porteur du sens plutôt que les termes « employé pour » lorsqu'ils souhaitent décrire un thème donné.

L'approche que nous résumons figure 5 permet d'exploiter de façon automatisée la liste de termes que nous identifions étape (ii) en nous appuyant sur un vocabulaire contrôlé riche. Pour chacun de ces termes, une première étape permet de vérifier si ce dernier n'apparaît pas en tant que concept dans l'ontologie. Dans ce cas, si ce terme existe dans le vocabulaire contrôlé, notre algorithme (figure 5) vérifie d'éventuelles approximations de sens avec les concepts de l'ontologie. Rappelons que nous souhaitons dans notre démarche utiliser l'ontologie de domaine définie pour accéder aux ressources du domaine. En effet, nous ne souhaitons pas dénaturer l'ontologie de domaine en modifiant sa structure et, par conséquent, nous limitons ici l'enrichissement à l'ajout de concepts et relations correspondantes au niveau des feuilles de l'ontologie. L'enrichissement consiste donc en l'ajout de concepts liés à des concepts de l'ontologie par des relations de spécialisation. L'idée est d'ajouter les termes subsumés i.e., les descendants dans l'arborescence.

La procédure d'enrichissement est présentée ci-dessous de façon synthétique.

Soient :

   *qEN* : un terme qualifiant non présent dans l'ontologie candidat à enrichir l'ontologie

   *liste_VedetteVocab* : la liste des termes du vocabulaire contrôlé utilisé

   *liste_termesCourant* : terme vedette courant du vocabulaire contrôlé utilisé et ses termes « employé pour » attachés

   *liste_conceptsFeuille* : 1 concept de l'ontologie et l'ensemble de ses concepts *fils*

1.   si *qEN* est présent dans *liste_VedetteVocab*

          *liste_termesCourant* ← liste de termes liés par la relation *employé_pour* à *qEN* le qualifiant qEN a un sens proche de celui de la liste liste_termesCourant

     a.   Si *liste_termesCourant et liste_conceptsFeuille* ont des termes équivalents

               i.   le qualifiant candidat *qEN* est alors attaché en tant que nouveau concept fils au concept de *liste_conceptsFeuille* qui possède le plus grand nombre d'équivalences à *liste_termesCourant* (la liste de termes de sens proche provenant du vocabulaire)

     b.   *sinon qEN* ne peut être ajouté à l'ontologie

**Figure 5. Algorithme général décrivant l'étape Enrichissement de l'ontologie**

Nous avons donc défini la méthodologie pour construire et enrichir une ontologie de domaine. Nous avons appliqué cette dernière dans le cadre d'un projet ANR GEONTO que nous présentons ci-après.

# 4. APPLICATION AU DOMAINE DE LA GEOGRAPHIE

Les ontologies de domaine s'attachent à décrire le vocabulaire particulier du domaine concerné. L'information géographique n'échappe pas à cette particularité. La multi dimensionnalité des objets géographiques confère des spécificités qui, en accord avec (Cullot et al, 2003), nous permettent de définir trois sortes d'ontologies géographiques :i) les ontologies cartographiques (au sens géographique) plus spécifiquement dédiées à la description des concepts qui caractérisent l'espace comme le point, la ligne, etc. Ces ontologies sont typiquement élaborées par de grands organismes de normalisation, par exemple, l'OpenGIS par le biais du langage GML (Geography Markup Language). ii) les ontologies de domaines géographiques comme une ontologie modélisant les concepts des données hydrauliques, ou encore les données des réseaux électriques, etc. Ce sont des ontologies « métier », développées par une communauté d'utilisateurs du domaine concerné. iii) les ontologies spatialisées (ou spatio-temporelles), qui sont des ontologies dont les concepts sont localisés dans l'espace. Cependant, les frontières entre ces domaines ne sont pas nettement délimitées. Nous en voulons pour preuve les travaux menés dans le cadre de la directive européenne INSPIRE[2] qui vise la mise en place d'une infrastructure d'information géographique au niveau européen.

De manière classique, les ontologies géographiques peuvent être utilisées pour l'exploration, ou encore pour l'extraction d'informations et au-delà pour l'interopération de SIG (Abadie &Munière, 2008), (Kokla, 2006), (Hess et al, 2007).

Notre travail se situe dans le domaine des ontologies spatialisées. Cette contribution présente l'intérêt d'une ontologie dans le cadre de l'annotation de documents textuels et plus particulièrement pour le typage automatisé des EN spatiales détectées dans des textes. Cette annotation s'appuie sur des ressources géographiques diverses (BD de l'IGN, *gazetteers* contributives) pour le typage et la validation d'EN spatiales candidates puis le calcul de géométries correspondantes. Ainsi, l'entité nommée *Artouste* aura une sémantique et une représentation spatiale différente selon la nature de l'élément géographique désigné. Qu'il s'agisse du lac, du pic ou de la vallée, les ressources invoquées et les stratégies de parcours de ces ressources seront différentes.

## 4.1  Contexte GEONTO
Le projet GEONTO est un projet ANR mené dans le cadre de l'édition 2007 du programme « Masse de Données et Connaissances ». Le consortium correspondant regroupe plusieurs laboratoires spécialisés, notamment dans le traitement de données cartographiques (COGIT[3]), la construction (IRIT- Université de

---

Toulouse) et l'alignement d'ontologies (LRI – Paris 11), ainsi que dans l'annotation et l'indexation automatisée d'informations géographiques dans des fonds documentaires textuels (LIUPPA – Université de Pau). Le COGIT dispose de bases de données géographiques hétérogènes et a pour objectif l'interopérabilité de ces bases. Pour cela, le projet prévoit de fournir une ontologie par base de données, et d'aligner les ontologies obtenues avec une ontologie de référence construite semi automatiquement par le COGIT.

## 4.2 Construction de l'ontologie géographique à partir de documents de description de ressources

Le COGIT dispose de plusieurs bases de données auxquelles sont associés des documents de spécification. Ces documents sont sémantiquement riches : les descriptions d'objets, de relations existant entre eux, de contraintes, de définitions, etc. sont exprimées à la fois par la structure matérielle du document et par le langage naturel. De plus, ces documents sont conformes à un XML schéma inspiré des normes ISO. L'expérimentation décrite ici porte sur la base de données BDTopo qui sert de référence pour la localisation de l'information relative aux problématiques d'aménagement, d'environnement ou d'urbanisme. Ses spécifications, disponibles au format WORD utilisant un style pour chaque type d'information, ont été automatiquement traduites en XML : c'est ce document qui servira à la construction de l'ontologie.

détaillée est donnée dans (Kamel & Aussenac, 2009). Nous présentons figure 6 un extrait de l'ontologie obtenue à partir des spécifications de la base de données BDTopo.

Le concept *Grotte*, fils du concept *Oronyme*, a pour concepts fils *Cave*, *Aven*, *Gouffre*, etc. La structure du document a également permis de caractériser des propriétés (A et D) telles que *a-pourNom*, ainsi que des relations sémantiques entre concepts (O) comme *a-pour-Importance* (Figure 6).

Un concept de l'ontologie a un label précisant sa position dans la hiérarchie du document. Ce choix a été motivé par le fait qu'un même terme peut se retrouver à différents niveaux dans le document. Une première solution est de créer un seul concept et de lui associer plusieurs concepts pères. Or, les spécifications donnent des définitions et des propriétés différentes dans chaque cas. Pour respecter la structure du document, nous avons choisi de concaténer le nom du concept courant à celui de ses concepts pères. Ceci permet de différencier les différents concepts, et de fournir par ailleurs une traçabilité de l'ontologie vers le document de spécification qui a servi à la construire.

De plus, les concepts sont documentés en définissant des propriétés : leur définition (*Definition*), les termes associés au concept (*Terme*), si le concept a été construit à partir de l'analyse de la structure ou de l'analyse du langage (*Origine*) ou s'il provient d'un attribut de classification tel que Nature (*Reference*). Ces informations sont utiles dans les phases de traçabilité ou d'alignement d'ontologies.

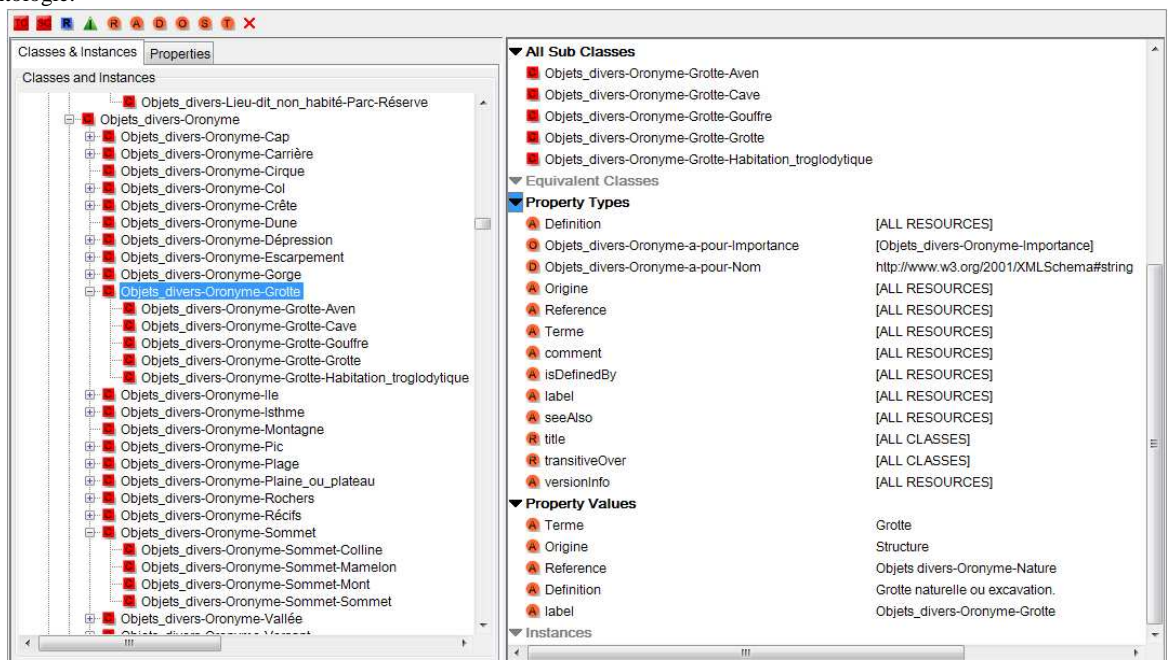

**Figure 6. Extrait de l'ontologie issue de la structure du document BDTopo**

### 4.2.1 Exploitation de la structure

Une étude systématique des balises et de leur organisation au sein du document de spécifications au format XML a permis de déterminer comment identifier concepts, relations conceptuelles et propriétés. Cette analyse ne présente pas de difficulté majeure car les balises véhiculent elles-mêmes leur sémantique et les relations découlent de la connaissance du domaine. Une description

### 4.2.2 Exploitation du texte rédigé

Les documents de spécification contiennent des textes très courts et très synthétiques, donc assez pauvres en matière d'expression de relations entre concepts. Le champ *définition* renferme néanmoins quelques expressions de la relation de méronymie ou de définition de propriétés. Pour analyser ce champ, une procédure d'annotation permet de caractériser les définitions par la séquence suivante :

> *{Concept} {Propriete}* ({M_PartieDe}) ? *{Terme} {Propriete}**

où *Concept* est un concept de l'ontologie, suivi éventuellement de 0 ou plusieurs propriétés (*Propriété*) caractérisées par un adjectif ou un complément du nom, éventuellement d'un marqueur linguistique spécifique de la relation de méronymie (*M_PartieDe*), d'un terme (*Terme*) obtenu à l'aide d'un extracteur de termes et éventuellement de 0 ou plusieurs propriétés.

L'enrichissement de l'ontologie est alors possible en exploitant ces définitions, selon l'algorithme suivant :

T : Terme, C : Concept, P : Propriété, ML : M_PartieDe

2. si seuls C et T sont présents, T devient un terme associé à C (relation de quasi-synonymie)

3. si seuls C, T et P sont présents, T est considéré comme un concept plus générique que C : C est relié à T par la relation *est-un*, et les propriétés sont associées à C.

   a. si T existe déjà dans l'ontologie, il n'y a pas de nouvelles relations créées
   
   b. si T n'existe pas dans l'ontologie, nous recherchons un terme plus spécifique ST inclus dans T (au sens lexical) qui serait un concept
   
      - si ST existe, T est relié à ST par la relation *est-un*
      
      - sinon T est créé comme un concept fils du concept Top

4. si C, T, ML et P sont présents, T est considéré comme un concept (reprendre l'algorithme à l'étape 2.) et C est relié à T par la relation *partie-de*

**Figure 7: Algorithme d'exploitation des définitions**

Les deux exemples présentés figures 8 et 9 illustrent la construction de bouts d'ontologie.

```
<pakage>
  <packageName> Voies de communication routière </packageName>
  <classe>
    <nom_classe> Tronçon de route </nom_classe>
    <définition> Portion de voie de communication
                 destinée aux automobiles </définition>
  </classe>
</pakage>
```

**Figure 8. Exemple de document structuré au format XML**

L'analyse de la structure du document établit une relation d'hyponymie entre les concepts *Tronçon de route* et *Voies de communication routière*. L'analyse du texte permet de créer le concept *Voie de communication* à partir du terme *voie de communication* comme un fils de *Top* et père de *Voie de communication routière* (car plus générique). Les concepts *Tronçon de route* et *Voie de communication* sont reliés par la relation *partie-de*, la propriété *destinée aux automobiles* est associée au concept *Tronçon de route*.

Voici figure 9 un extrait de document fourni par le COGIT décrivant l'oronyme *Grottes*.

```
<pakage>
  <packageName> Objets Divers </packageName>
  <classe>
    <nom_classe> Oronyme </nom_classe>
    <définition> grotte naturelle ou excavation </définition>
    <valueName> Grotte </valueName>
  </classe>
</pakage>
```

**Figure 9. Exemple de document structuré au format XML**

L'analyse de la structure du document établit une relation d'hyponymie entre les concepts *Objets Divers, Oronyme et Grotte*. L'analyse du texte permet d'associer les termes *Grotte naturelle* et *Excavation* au concept Grotte.

Nous montrons dans le paragraphe suivant comment extraire et analyser des termes associés à des EN spatiales dans le but d'enrichir l'ontologie obtenue.

## 4.3 Extraction et analyse de termes associés à des EN spatiales dans des textes grand public

### 4.3.1 Description de la chaine de traitement
La chaine de traitement que nous proposons (figure 10) s'appuie sur une démarche de « tokenisation » classique (1).

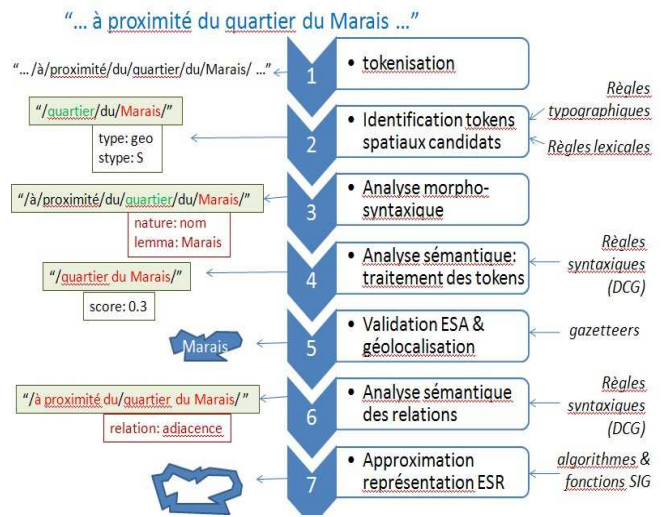

**Figure 10. Chaine de traitement d'information spatiale dans des documents textuels.**

Nous adoptons une démarche de lecture « active » qui consiste à marquer rapidement des EN spatiales candidates puis à appliquer les étapes suivantes de l'analyse à ces EN uniquement. Un marqueur de *token* spatial candidat (2) utilise des règles lexicales (lexiques d'introducteurs d'information spatiale) et typographiques (majuscule en début de token). Puis, un analyseur morpho-syntaxique associe un lemme et une nature à chaque token spatial candidat (i.e. *Marais*, nom). Un analyseur sémantique (4) et (6) associé à des règles de grammaire DCG (Definite Clause Grammar) exprimées en Prolog qualifie des entités spatiales absolues (ESA) et des entités spatiales relatives (ESR). Une ESA est une EN simple : *le quartier du Marais, la rue Emile Zola, la ville de Pau, le pic d'Ossau, la vallée d'Ossau,* … Une ESR est une EN complexe définie à partir d'une autre EN : *au cœur du quartier du Marais, dans la périphérie de la ville de Pau, au sud de la vallée d'Ossau,* … Les ESA sont validées et géolocalisées (5) à l'aide de ressources externes ou de *gazetteers* contributives internes. Les ESR sont construites à partir des ESA ainsi détectées. Les relations spatiales correspondantes sont interprétées et des représentations géométriques approximées sont calculées. Cette chaine de traitement spatiale est détaillée dans (Gaio et al., 2008) et (Loustau et al., 2008).

### 4.3.2 Analyse de termes associés à des EN spatiales

À l'aide de cette chaîne de traitement, nous avons analysé 14 livres (récits de voyage dans les Pyrénées) afin d'obtenir deux ensembles de termes (ceux associés à des toponymes validés par des ressources géographiques et ceux associés à des toponymes candidats non validés) présentés précédemment et illustrés en figure 11.

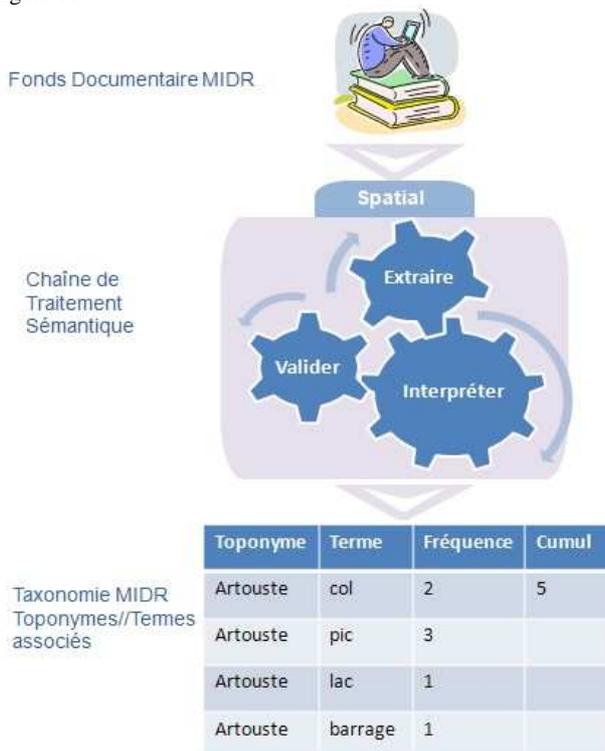

**Figure 11. Extraction de termes associés à des EN spatiales.**

Dans le §4.5, le lecteur pourra apprécier notre étude quantitative issue de cette analyse.

Nous avons également mené une étude qualitative des termes associés. Si l'on considère l'extrait de l'ontologie géographique générée (figure 12) et la liste des termes communs associés à des EN spatiales dans des récits de voyage, nous observons que 50% des termes sont communs à des concepts du niveau Nature dans l'ontologie et 50% sont communs à des concepts de niveau Sous-nature. Ce dernier niveau étant le résultat du traitement automatisé des spécifications textuelles des BD de l'IGN présenté §4.2.

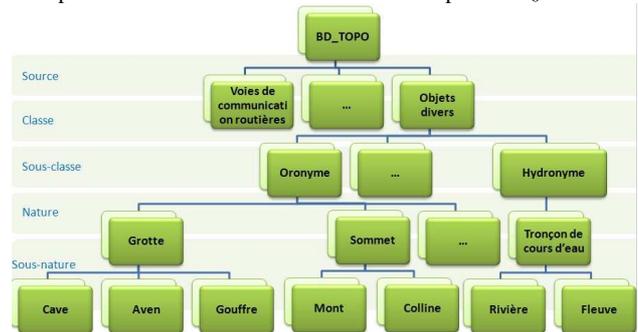

**Figure 12. Extrait de l'ontologie géographique.**

Nous observons qu'un grand nombre des termes distincts associés à des toponymes dans des récits de voyages pourraient venir enrichir les concepts de niveau Sous-nature. Prenons l'exemple des termes *abîme, antre, caverne* qui reviennent régulièrement dans le corpus analysé ; il apparait clairement qu'ils pourraient enrichir le niveau Sous-nature de l'ontologie par autant de nouveaux concepts (cf. figure 12 le sous arbre relatif au concept Grottes). Nous envisageons d'utiliser l'échantillon de termes distincts ainsi extrait et de l'associer au potentiel du thésaurus RAMEAU[4] en vue de l'enrichissement de l'ontologie géographique. RAMEAU (Répertoire d'autorité-matière encyclopédique et alphabétique unifié) est le langage d'indexation matière utilisé, en France, par la Bibliothèque nationale de France, les bibliothèques universitaires, de nombreuses bibliothèques de lecture publique ou de recherche ainsi que plusieurs organismes privés. Il se compose d'un vocabulaire de termes reliés entre eux et d'une syntaxe indiquant les règles de construction des vedettes-matière à l'indexation afin d'en assurer le bon usage. RAMEAU est une source riche en connaissances avec plus de 400000 termes de domaines variés ; il couvre l'ensemble des disciplines scientifiques et contient aussi les termes traitant des loisirs, des arts, etc.

Si l'on prend l'exemple du toponyme Crabioules (figure 13) extrait de notre échantillon et associé à 8 termes dans ces mêmes textes, nous pouvons remarquer (figure 14) que *abîme* a pour terme vedette *Grottes* dans RAMEAU et que les termes « employé_pour » *Grottes* sont *abîme, antre, aven, caverne, gouffre,* etc.

| Crabioules | | | |
|---|---|:---:|:---:|
| **Occurrences** | **Terme associé** | **Ontology Géographique** | **Thesaurus RAMEAU** |
| 1 | abîme | | ✓ |
| 2 | col | ✓ | ✓ |
| 1 | corniche | | ✓ |
| 1 | crête | ✓ | ✓ |
| 1 | mont | ✓ | ✓ |
| 1 | promenade | | ✓ |
| 1 | route | ✓ | ✓ |
| 1 | sommet | ✓ | ✓ |

**Figure 13. Liste des termes associés au toponyme Crabioules**

---

[4] http://rameau.bnf.fr/informations/rameauenbref.htm

« Crabioules » est qualifié dans l'échantillon de textes analysés par divers termes qui peuvent renvoyer à des représentations spatiales différentes du toponyme. L'ontologie géographique présentée apporte des éléments de réponses en permettant notamment d'identifier une représentation spatiale pour les entités spatiales *Col de Crabioules* et *Mont de Crabioules* (cf. *Col* et *Mont* dans figure 13). Cependant, un certain nombre de termes pose encore problème et nous souhaitons exploiter la structure du thésaurus RAMEAU pour enrichir cette ontologie et lever des ambiguïtés.

L'ontologie géographique offre donc un premier élément de réponse à la nécessité de typage des entités géographiques détectées dans des textes. Le potentiel de notre échantillon de récits de voyages associé au thésaurus RAMEAU est relativement important. Dans la partie suivante, nous proposons d'utiliser les termes associés à des toponymes dans des textes grand public afin d'enrichir l'ontologie géographique. La méthode décrite ci-après s'appuie sur le thésaurus RAMEAU et le recoupement de sous-arbres de l'ontologie avec des sous-arbres du thésaurus RAMEAU.

## 4.4 Enrichissement de l'ontologie géographique

Prenons l'exemple du terme *Grottes* qui est défini en tant que vedette dans RAMEAU et dont la notice descriptive est présentée figure 14.

**Grottes** [+ subd. géogr.]

*Vedette matière nom commun . S'emploie en tête de vedette*

**<Employé pour :**
Abîmes
Antres
Avens
Cavernes *Ancienne vedette*
Cavernes préhistoriques
Cavités souterraines
Gouffres
Grottes ornées
Grottes préhistoriques
Préhistoire – Grottes
Spélonques

**<<Terme(s) générique(s) :**
Habitat préhistorique
Relief (géographie)
Zones souterraines

**Figure 14. Exemple de notice descriptive dans RAMEAU décrivant le terme *Grottes***

Grottes possèdent pour termes « employés pour » les termes *abîmes*, *antres*, *gouffres*, etc. Un catalogueur qui travaille sur la description d'un document image représentant les *Abîmes de Crabioules* devra utiliser le terme vedette *Grottes* plutôt que *Abîmes* dans la notice descriptive du document.

### 4.4.1 Association d'un terme à un concept

L'objectif est ici d'enrichir l'ontologie géographique par les termes qualifiants les entités spatiales dans un échantillon représentatif du corpus documentaire traité afin de leur faire correspondre une représentation spatiale adéquate.

Reprenons l'exemple de l'entité spatiale *les Abîmes de Crabioules*. Le terme *Abîmes* est alors identifié par notre chaîne de traitement dans l'échantillon des textes traités comme un qualifiant d'une entité spatiale. L'ontologie géographique ne permet pas d'identifier la représentation spatiale adéquate car le concept *Abîmes* n'existe pas. Pour enrichir l'ontologie géographique, nous proposons d'appliquer la méthodologie présentée figure 7 dans laquelle nous exploitons le thésaurus RAMEAU afin de vérifier si ces termes, n'apparaissant pas dans l'ontologie, existent dans le thésaurus et le cas échéant s'il est possible d'identifier des approximations de sens avec des concepts de l'ontologie.

Si nous reprenons l'exemple *Abîmes* inconnu jusque là dans l'ontologie, nous vérifions si ce terme est présent dans le thésaurus et lorsque c'est le cas (la figure 14 montre bien que c'est le cas ici), nous récupérons la vedette identifié ainsi que l'ensemble de ses termes « employé pour ». Cela indique que le qualifiant candidat à enrichir l'ontologie a un sens proche à l'ensemble de ces termes sélectionnés dans le thésaurus. Nous regroupons ensuite par classes les concepts de l'ontologie. Pour cela, nous créons une classe lorsque nous identifions un concept possédant des concepts fils feuilles (de bas niveau). Le concept père est alors le représentant de la classe et ses fils feuilles, reliés par les relations « is_a » ou « part_of » sont ajoutés à la classe. Nous comparons chaque classe de concepts à l'ensemble de termes identifié dans le thésaurus RAMEAU qui ont un sens proche au qualifiant candidat. Si des équivalences sont identifiées. Le qualifiant candidat, possédant le plus grand nombre d'équivalences à la liste de termes de sens proche provenant de RAMEAU, devient alors nouveau concept de l'ontologie en tant que feuille attachée par la relation « is_a » au concept représentant de la classe identifiée comme la plus proche.

Pour le candidat *Abîmes*, nous identifions trois équivalences entre les concepts de l'ontologie géographique sous le concept *Grottes* (figure 12) et la liste de termes provenant de RAMEAU liés à la vedette *Grottes* (figure 14). En effet, Grottes, Avens et Antres sont présents dans les deux ensembles et cela nous permet de créer un nouveau concept *Abîmes* en tant que *fils* au concept *Grottes* (figure 15).

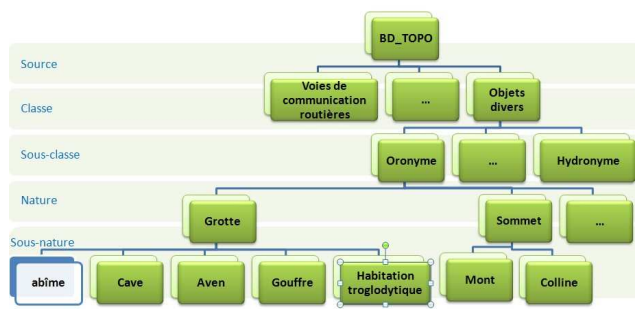

**Figure 15 : Extrait de l'ontologie géographique enrichie**

La méthodologie proposée permet ainsi d'enrichir l'ontologie en ajoutant de nouveaux concepts afin de proposer une couverture sémantique du domaine *géographie* plus importante. L'ontologie ainsi mise en place donne alors la possibilité de lever des ambiguïtés dans la phase d'identification de la représentation spatiale adaptée à l'entité spatiale identifiée.

Il reste cependant des termes pour lesquels RAMEAU ne nous permet pas d'apporter de réponse. De plus, certains des termes qui enrichissent l'ontologie géographique peuvent engendrer des contres sens qui peuvent amener à générer des résultats erronés. En effet, les termes *glacier* et *gorges* notamment ont un double sens (géographique et autre), et ils sont chacun présents deux fois, sous des écritures différentes, dans RAMEAU. Cela implique de travailler plus en détails sur le contexte de chaque terme dans RAMEAU afin d'identifier la bonne vedette à exploiter.

Nous travaillons actuellement sur ces points en étudiant les ressources (EuroWordNet, Larrousse) qui pourraient nous permettre, de la même façon que RAMEAU, de clarifier ces ambiguïtés. Nous travaillons également sur l'intégration, au sein de notre chaine de traitement, de l'ontologie géographique développée afin de valider notre approche sur un fonds documentaire complet. Une analyse détaillée doit en effet être menée sur un nombre conséquent de documents pour comparer les résultats et notamment le nombre d'entités spatiales pour lesquelles l'ontologie géographique nous permet d'identifier de façon automatique une représentation spatiale appropriée.

## 4.5 Evaluation

L'analyse des termes associés aux entités nommées et l'utilisation des données des BD de l'IGN nous permet de typer automatiquement 5% des informations spatiales annotées (15% du nombre total d'occurrences). Ce taux passe à 10% de typage automatique des informations spatiales (33% du nombre total d'occurrences) si l'on utilise l'ontologie géographique générée dans le cadre du projet GEONTO. Nous obtenons 50% (75% du nombre total d'occurrences) de typage automatique des informations spatiales après l'enrichissement de cette ontologie à partir d'une étude combinée d'échantillons de textes grand public (récits de voyage) et du thésaurus RAMEAU.

A l'issue de l'étude quantitative des termes associés, nous avons obtenu :

- 15560 occurrences (9625 distincts) d'EN spatiales candidates et 15560 occurrences (2388 distincts) de termes associés (figure 16a) ;

- 8773 occurrences (4705 distincts) d'EN spatiales validées et 8773 occurrences (1526 distincts) de termes associés (figure 16b).

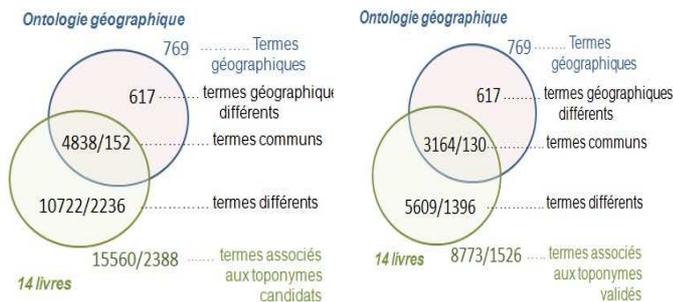

**Figure 16a. Toponymes candidats & termes associés**

**Figure 16b. Toponymes validés & termes associés**

Nous obtenons des termes communs à ceux de l'ontologie géographique :

- 4838 occurrences de termes associés à des toponymes candidats sont communs, soient 152 termes communs distincts (figure 16a) ;

- 3164 occurrences de termes associés à des toponymes validés sont communs, soient 130 termes communs distincts (figure 16b).

Nous obtenons des termes différents de ceux de l'ontologie géographique :

- 10722 occurrences de termes associés à des toponymes candidats sont différents, soient 2236 termes différents distincts (figure 16a) ;

- 5609 occurrences de termes associés à des toponymes validés sont différents, soient 1396 termes différents distincts (figure 16b).

Ces premiers éléments statistiques ont été améliorés via l'intégration du thésaurus RAMEAU.

Ainsi, nous avons estimé l'enrichissement de l'ontologie via RAMEAU comme présenté figure 17.

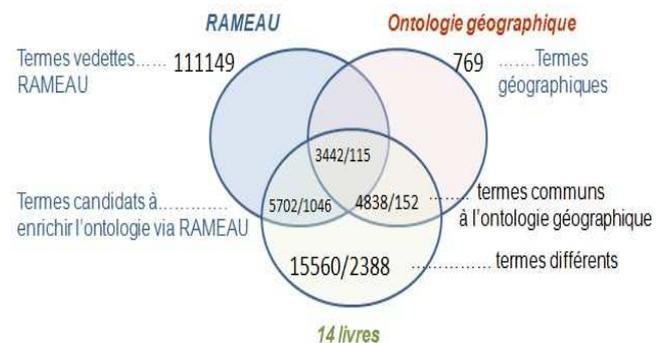

**Figure 17. toponymes candidats & termes associés.**
Et donc 1046 termes RAMEAU sont candidats à l'enrichissement de notre ontologie. Nos travaux actuels affinent cette dernière statistique afin de détecter les termes RAMEAU porteur d'un sens géographique.

## 5. CONCLUSION ET PERSPECTIVES

Ce travail présente une méthode pour la construction d'ontologie couvrant au mieux un domaine sémantique. Dans un premier temps, cette méthode s'appuie sur le contenu et la structure de documents de description de ressources pour la création automatique d'ontologie de domaine. L'étape suivante consiste à enrichir l'ontologie obtenue à partir d'une analyse de textes et d'un vocabulaire contrôlé du même domaine.

L'ontologie obtenue peut être utilisée dans diverses applications. Nous avons expérimenté la création puis l'utilisation d'une ontologie géographique dans le cadre de corpus textuels composés de récits de voyages dans les Pyrénées. Nous avons montré que l'utilisation de l'ontologie géographique dans la chaine de traitement pour l'annotation automatique d'EN spatiales dans un corpus qui décrit un territoire (récits de voyage, journaux, etc.) a permis de diminuer de plus de moitié les ambiguïtés pouvant apparaître lors de la validation et du calcul de représentations associées aux EN détectées. L'ontologie géographique générée offre ainsi un moyen efficace d'optimiser l'accès à des ressources grand public de type récits de voyages.

Nous travaillons actuellement sur l'affinement des résultats obtenus afin de détecter les termes RAMEAU porteur d'un sens géographique. Nous cherchons également à identifier les termes pouvant engendrer des contres sens qui peuvent amener à générer des résultats erronés. Cela implique de travailler plus en détails sur le contexte de chaque terme dans le vocabulaire contrôlé utilisé afin d'identifier la bonne vedette à exploiter. Nous souhaitons également mener à court terme une analyse détaillée sur un nombre conséquent de documents pour comparer les résultats et notamment le nombre d'entités spatiales pour lesquelles l'ontologie géographique nous permet d'identifier de façon automatique une représentation spatiale appropriée.

# 6. ACKNOWLEDGMENTS